\documentclass[10pt,a4paper]{article}
\usepackage[top=2cm,left=7.6cm,footskip=1cm]{geometry}

\usepackage{amsmath,amssymb}

\usepackage{changepage}

\usepackage[utf8x]{inputenc}

\usepackage{textcomp,marvosym}

\usepackage{cite}

\usepackage{nameref,hyperref}

\usepackage[right]{lineno}

\usepackage{microtype}
\DisableLigatures[f]{encoding = *, family = * }

\usepackage[table]{xcolor}

\usepackage{array}

\newcolumntype{+}{!{\vrule width 2pt}}

\newlength\savedwidth

\raggedright
\usepackage[aboveskip=1pt,labelfont=bf,labelsep=period,justification=raggedright,singlelinecheck=off]{caption}

\makeatletter
\renewcommand{\@biblabel}[1]{\quad#1.}
\makeatother

\usepackage{lastpage,fancyhdr,graphicx}
\usepackage{epstopdf}
\usepackage{multirow}
\usepackage{subcaption}

\newcommand{\beginsupplement}{%
        \setcounter{table}{0}
        \renewcommand{\thetable}{S\arabic{table}}%
        \setcounter{figure}{0}
        \renewcommand{\thefigure}{S\arabic{figure}}%
     }

\begin{document}
\vspace*{0.2in}

\begin{flushleft}
{\Large
\textbf\newline{A network-based citation indicator of scientific performance}
}
\newline
\\
Christian Schulz\textsuperscript{1*}, Brian Uzzi\textsuperscript{2,3}, Dirk Helbing\textsuperscript{1}, and Olivia Woolley-Meza\textsuperscript{1}
\\
\bigskip
\textbf{1} Computational Social Science, ETH Zurich, Zurich, Switzerland
\\
\textbf{2} Kellogg School of Management, Northwestern University, Evanston, IL, USA
\\
\textbf{3} Northwestern Institute on Complex Systems, Northwestern University, Evanston, IL, USA
\\
\bigskip

* cschulz@ethz.ch

\end{flushleft}
\section*{Abstract}
Scientists are embedded in social and information networks that influence and are influenced by the quality of their scientific work, its impact, and the recognition they receive.   Here we quantify the systematic relationship between a scientist's position in the network of scientific collaborations and the citations they receive.    As expected, we find that authors closer to others in this network are, on average, more highly cited than those further away from others.           We construct a novel indicator, the $s$-index, that explicitly captures performance linked to network position along two complimentary dimensions: performance expected due to network position and performance relative to this position. The basis of our approach is to represent an author's network position through their distribution of distances to other authors.     The $s$-index then ranks (1) the \textit{citation potential} of an individual's network position relative to all other authors, and (2) the citations they accrue relative to authors that have a comparable network position.    Characterizing scientists through these two complimentary dimensions can be used to make more informed evaluations in a networked environment. For example, it can identify individuals that play an important role in diffusing scientific ideas. It also sheds a new light on central debates in the \textit{Science of Science}, namely the impact of author teams and comparisons of impact across scientific fields.   
\section*{Introduction}

There is a rich literature on measuring scientific success \cite{garfield1955impact,de1965networks,hirsch2005index,wang2013quantifying,fortunato2018science}. Most success indicators are based on citation counts (for an overview see \cite{waltman2016review}).    Citation accrual is a complex process, which is influenced by the diffusion of ideas through the interpersonal networks scientists are embedded in \cite{petersen2015quantifying}.     For example, consider networks of collaboration \cite{newman2001structure,barabasi2002evolution,borner2004simultaneous}. A scientist is more likely to know, and therefore cite, their own work than the equally relevant work of other authors.   Through the same information mechanism, scientists are more likely to become familiar with and cite the work of their co-authors, their respective co-authors, and so on. These dynamics may result in a system where connections to other scientists can lead to more citations. Furthermore, feedback can magnify this process as citations increase visibility \cite{merton1968matthew,price1976general,fowler2007does,petersen2014reputation}. Currently, the relationship between citation counts, network positions, and the diffusion of ideas in science is not well understood. This suggests that performance measures that incorporate network information may help broaden our understanding of the spread of scientific ideas and the allocation of credit.  

 Some attempts have been made to incorporate network information in citation indicators by weighting citations relative to where in a scientist's network they come from. Most simply, self-citations can be discounted \cite{aksnes2003macro}. Going a step further, a $c$-index \cite{bras2011bibliometric} (analogous to the $h$-index) includes only the $c$ citations that come from at least a distance $c$ from an author. However, these approaches do not have a theoretical or empirical basis. Such discounting of citations based on social distance can be problematic because network position is not independent from the quality of an author's work. Good work can lead to a better network position and the converse is also possible. Additionally, building a good network position is an important scientific skill in itself, facilitating the diffusion of ideas.             

                  Here we present a new approach that avoids these problems. Namely, we propose a novel author-level citation indicator, the $s$-index, that characterizes performance through two complimentary dimensions: 1) the performance associated with an individual's network position and 2) the performance that is network independent.    Specifically, we first build a quantitative, data-driven model of citation impact as a function of network position. With this baseline we then give the author a two-dimensional score: 1) the rank of the citation potential of an individual's network position compared to all scientists, 2) the rank of their realized citations compared to others who have network positions with a comparable citation potential.     Beyond its use in evaluating individuals in a networked world, this indicator can shed a new light on important questions about scientific impact in general. For example, here we investigate to what degree differences in network position can account for the different impact of scientific fields, or the higher impact of multi-author publications \cite{wuchty2007increasing}.

\begin{figure}
\begin{adjustwidth}{-6.0cm}{-0mm}
\includegraphics[width=18.0cm]{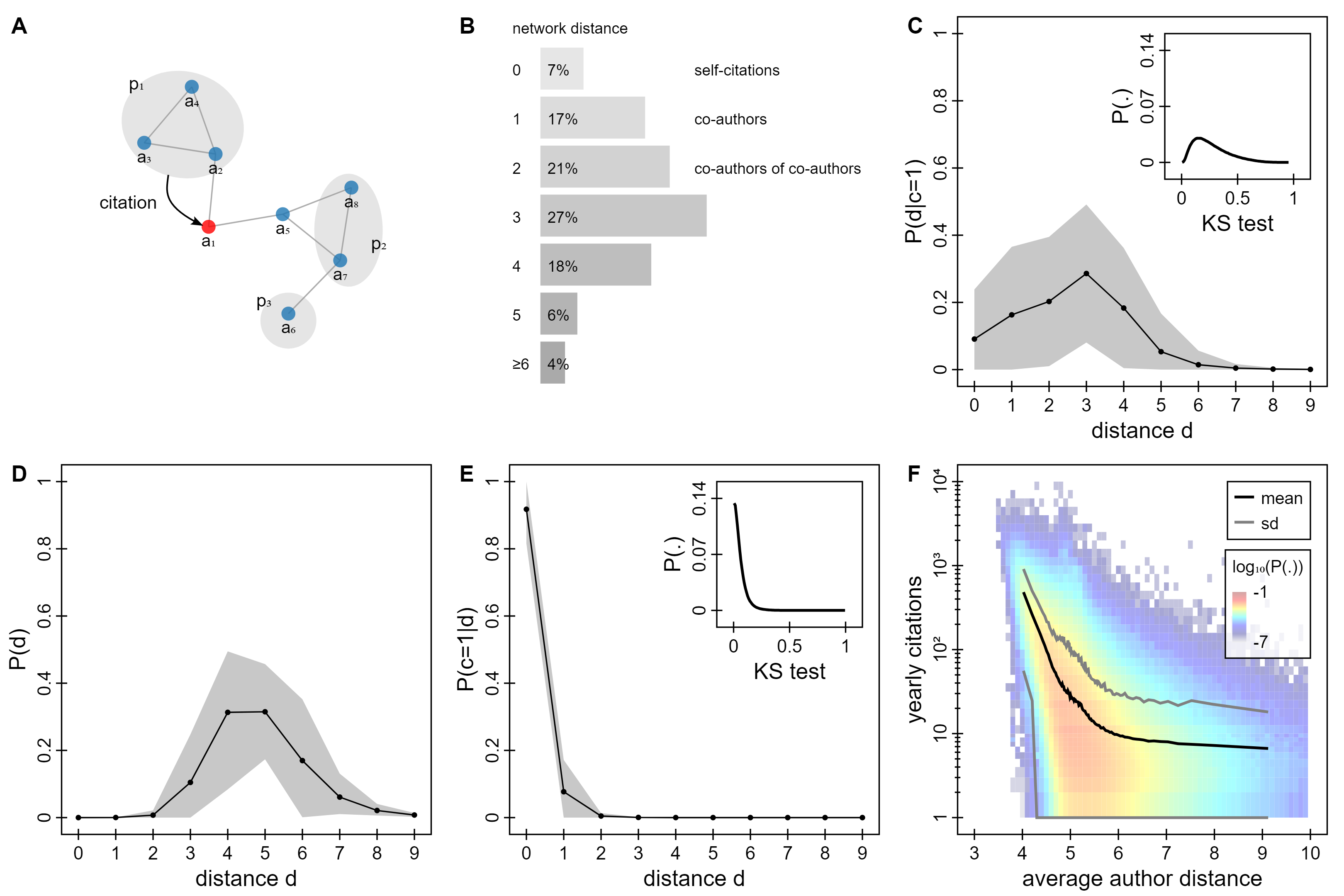}
\end{adjustwidth}
\caption[Citation is more likely from authors closer in the co-author network]{\textbf{ Citation is more likely from authors closer in the co-author network.} \textbf{(A)} Each citation an author receives can be mapped to a distance on this network. Specifically, given a focal author, e.g. $a_1$, the distance of a citation source is computed by considering the closest author in the set of authors of the citing paper.  \textbf{(B)} Distance of citations to cited authors in 2010. \textbf{(C-E)} Author-level probability mass functions of distances and citations on the co-author network: (C) Probability distribution of distances between the set of authors of the citing article and the cited author. (D) Probability distribution of distances to all articles published. (E) Probability that any paper at distance $d$ from a focal author will cite her. The mean of all authors is shown by the marker and the gray area spans 95\% of the data. We consider 120,000 authors with at least 1,000 citations received between 2000-2009. Insets (C and E) show the distribution of the KS statistic of pairwise comparison of the distributions of randomly drawn authors. Distances from which citations are accrued (C) vary strongly between authors. There is much greater similarity between authors when we control for their network position, i.e. dividing through $P_i(d)$ to obtain the probability an author will receive a citation from an author at a given distance (E). \textbf{(F)} The average distance from an author to the rest of the network vs. the number of citations they receive in a year.   Authors that are at small average distances receive many more citations. We observe a clear functional relationship between the mean citations received given an average distance.   }
\label{neteffect}
\end{figure}

\begin{figure}
\begin{adjustwidth}{0.0cm}{-0mm}
\includegraphics[width=12.0cm]{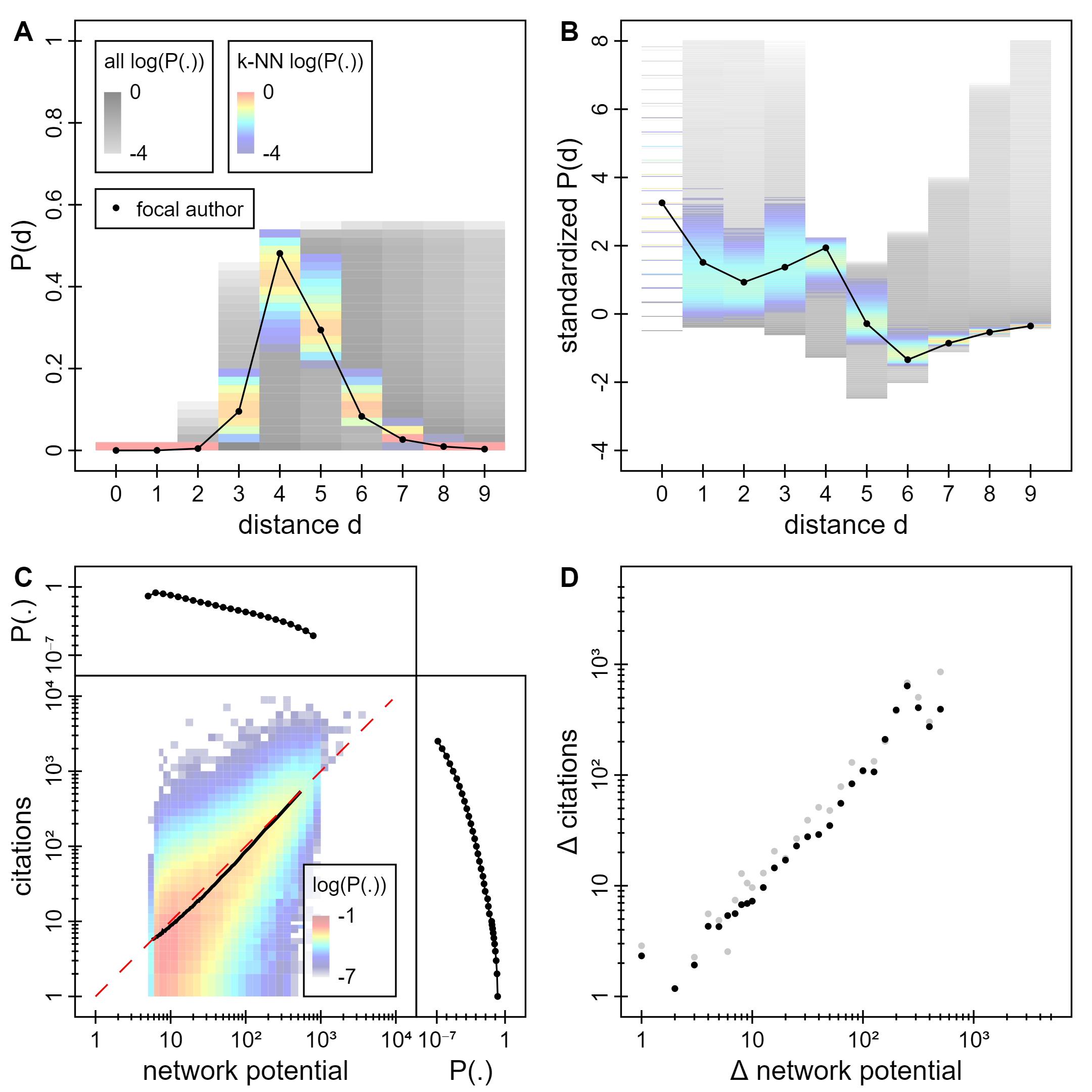}
\end{adjustwidth}
\caption[Revealing the network potential]{\textbf{Revealing the \textit{network potential}.}  (Data: all authors active in 2010, citations received in 2010.) For each author, the network potential is the average citation performance of authors with the most similar network positions according to their \textit{network profiles}.  \textbf{(A)} The \textit{network profile} (distance distribution $P_i(d)$) for a randomly chosen author and the density (color-scale) of probability mass of its 1,000 nearest neighbors and all authors (in gray-scale). \textbf{(B)} Same as A, but with each element of $P_i(d)$ standardized ($\mu=0$ and $\sigma=1$). Authors are represented as a point in the 10-dimensional vector of standardized $P_i(d)$s. To find their $k$-NN we consider the Euclidean distance in this space. The $k$-NN exhibit similar distance distributions despite the strong variability across the entire set of authors. \textbf{(C)} Joint and marginal distributions of the actual citations and the \textit{network potential}. The empirical citations are more heterogeneous than the potential, indicating, as expected, that factors other than network position also influence citations received. However, the mean empirical citations conditional on a given \textit{network potential} (black solid line), closely match the potential. \textbf{(D)} Effect of an increase of network potential on the change of number of received citations in the following year. A matching method \cite{stuart2010matching} finds for each author experiencing an increase in network potential the most similar one without a positive change according to other observable variables. The average change of citations (grey dots) is then subtracted by the average outcome of the matched authors and the result is shown by black dots. See also SI Fig.~\ref{matching} for the smaller magnitude effect in the other direction.}
\label{volumeind}
\end{figure}

\section*{Results}

\subsection*{Co-authorship network and citation dynamics}

What does an author's professional network reveal about how their work is cited? To answer this question we consider the interpersonal connections between scientists that form through co-authorships. We build yearly co-authorship networks $\textbf A$$(y)$, where the nodes are publishing authors. Two authors $i$ and $j$ are connected by an unweighted edge $a_{ij}$ if they have co-authored at least one paper published before the year $y$. Authors are considered active if they have published at least once in the previous 5 years.   In 2010, the giant network component consists of 82.1\% of the 4.26 million authors.  

 The co-authorship network induces paths between authors, which can be represented by an ordered set of links $a_{ij}$. The length of a path is the number of links that are traversed from the beginning to the endpoint of the path. The distance $d_{ij}$ between two authors $i$ and $j$ is the length of the shortest path between them. Similarly,   we denote with $d_{ip}$ the distance between an author $i$ and a paper $p$. As shown in Fig.~\ref{neteffect} A, this is simply the length of the shortest path between $i$ and any of the authors $j$ of paper $p$ (paper $p$ is simply a set of authors $j\in p$):

\begin{equation}
d_{ip}=\min\sum_{k,m \in \Gamma_{ip}} a_{km}, 
\end{equation}
where $\Gamma_{ip}=\cup_{j\in p} \Gamma_{ij}$ and $\Gamma_{ij}$ is the set of paths between $i$ and an author $j$ of paper $p$. Using the minimal distance to the set of authors of paper $p$ is in line with the notion of a self-citation, where only one of the authors of the citing publication needs to be the cited author.

We clearly see that an author is disproportionately likely to be cited by authors at short distances. In 2010 the average distance between an author and a citing paper was only 2.5 (Fig.~\ref{neteffect} B), while the average distance to all papers was 5.7.     

 The bias towards citation of proximate authors is even clearer when we compare to the baseline distribution of distance to any paper. Specifically, let us compare the distribution of the probability $P_i(d=x|c=1)$ that a paper citing an author $i$ (in year $y$) is at distance $x$, with the baseline (unconditional) distance distribution $P_i(d=x)$ from author $i$ to all papers (published in year $y$).   As expected, the distribution conditioned on citation is skewed towards shorter distances compared to the baseline (see Fig.~\ref{neteffect} C-D).    The probability that a publication at a given distance will cite an author can be found controlling for this baseline, using Bayes' formula,   $P_i(c=1|d=x) = P_i(c=1) P_i(d=x|c=1) / P_i(d=x)$. Here, the probability of receiving a citation $P_i(c=1)$ is chosen to ensure proper normalization. The results are shown in Fig.~\ref{neteffect} E. We find, for example, that an article published by the focal author has a probability of about 0.9 of citing the focal author (in this case a self-citation). An article published by a former co-author has a smaller, but still significant, probability of about 0.1.

 Interestingly, whereas there is significant author level variation in the distance distributions, the probability of citation given a distance, $P_i(c=1|d)$, exhibits much stronger regularity. A more rigorous comparison can be obtained through a pair-wise Kolmogorov-Smirnov (KS) test of the individual distributions (see Fig.~\ref{neteffect} C,E insets). The pairwise comparison of the author level distribution of $P_i(c=1|d)$ and $P_i(d|c=1)$ yields an average KS distance of 0.056 and 0.262 respectively. This indicates that the author's position in the network accounts for important variation in the source of received citations.  

 Summarizing the full distribution $P_i(d)$ using the average distance from an author to the rest of the network, commonly referred to as closeness centrality, we find that even this very coarse measure of network position exhibits a strong relationship with $C_i(y)$, the total citations an author $i$ receives over a specific year (see Fig.~\ref{neteffect} F). The average citations exhibit a clear functional relationship with distances, as they decrease with distance   (see SI for a more detailed analysis of the systematic patterns of variation in citations around this average behavior). This is consistent with independent prior work indicating that network centrality measures predict citation success \cite{sarigol2014predicting}. Beyond the average dynamics, this analysis also reveals that, strikingly, individual authors who are not well connected rarely reach high citation levels.

\subsection*{Quantifying the citation potential of network position}

 In order to build our indicator we need a model for the yearly citations that we should expect from a scientist given only information about their position in the co-author network.      There are many plausible models for the \textit{network potential}. We take a data-driven approach, making use of a large dataset of millions of scientists, over varied network positions, to extract the empirical regularity in the citations received given authors' network positions.   Thus, we use the notion that the network position of author $i$ is captured by the distribution of distances to the sets of authors of newly published articles, $P_i(d)$.    The \textit{network potential} of an individual in a given year $y$ is then determined by considering the performance of others in similar positions. Specifically, we find the mean total citations accrued in year $y$, over every author in the set of 1,000 authors with the most similar distance distribution to $i$ (see Fig.~\ref{volumeind} A-B and Materials and Methods for details). Characterizing an individual's network position in this way yields better predictions of an author's yearly accrued citations than other methods, for example, using the degree centrality (number of co-authors) and closeness centrality (see SI for more details). As desired, we can see in Fig.~\ref{volumeind} C   that the mean empirical citations, conditional on the \textit{network potential}, closely match the \textit{network potential}. For any given \textit{network potential}, there is significant idiosyncratic variation in the performance of authors around the expectation. The \textit{network potential} is more homogeneously distributed than the empirical citations, precisely because the former only captures the variability that is captured by network position.

\subsection*{Constructing the $s$-index}

\begin{figure}
\begin{adjustwidth}{0.0cm}{-0mm}
\includegraphics[width=12.0cm]{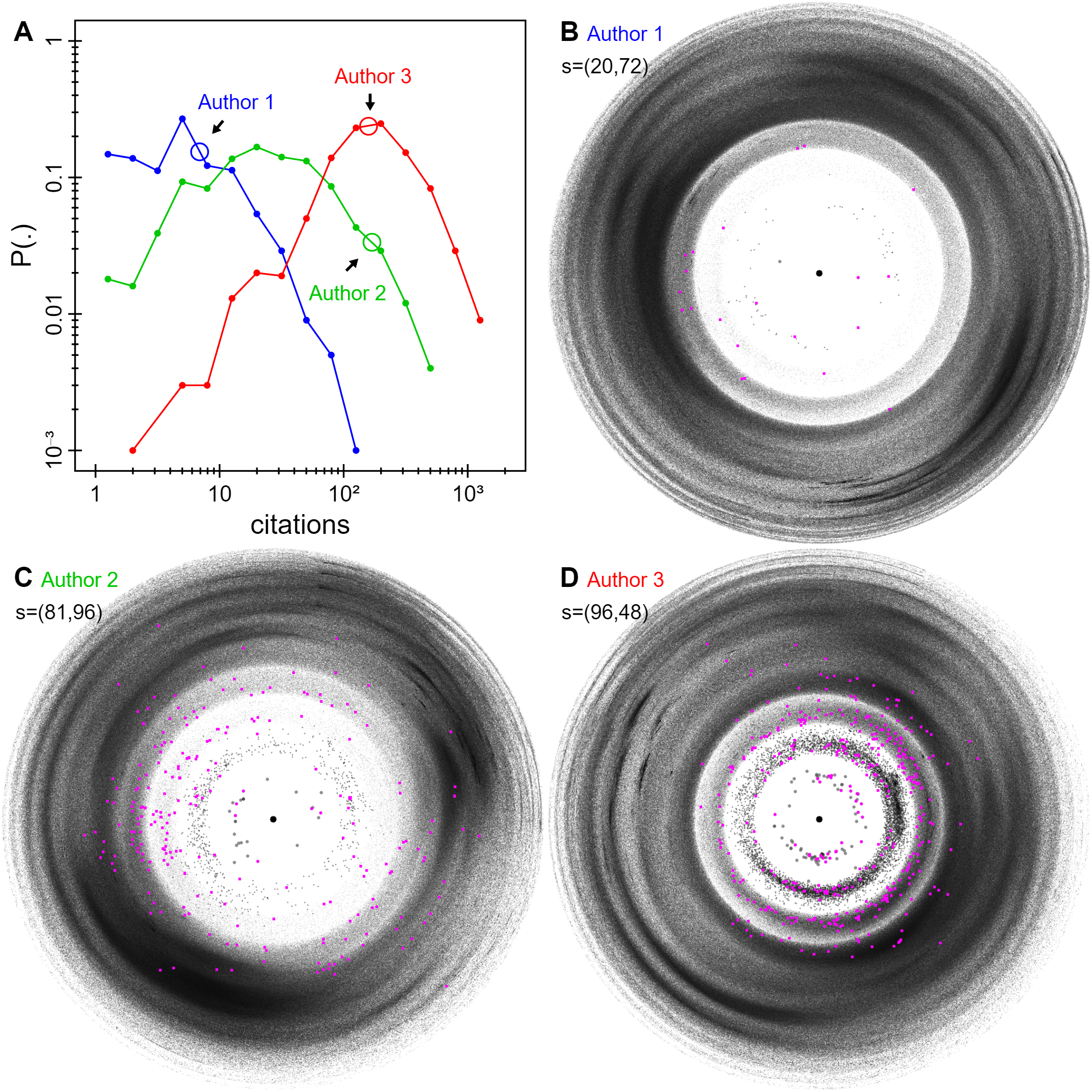}
\end{adjustwidth}
\caption[$s$-index examples]{\textbf{$s$-index examples.}   \textbf{(A)} We use data from 2016 to illustrate the citation count distribution of the 1,000 nearest neighbors of three example researchers. The average value of a curve is the network potential and the rank within this distribution forms the $s_{\operatorname{pers}}$ value. With seven citations for the year 2016, researcher \textit{1} achieves an $s_{\operatorname{pers}}$ of 72 among the comparison group for a low network potential. Author \textit{2} has an outstanding $s_{\operatorname{pers}}$ performance, even given the above average network position. Finally, researcher \textit{3}, who reached a similar citation level as researcher \textit{2}, scores exceptionally for $s_{pos}$, but only average for $s_{pers}$. \textbf{(B-D)} Collaboration network from the perspective of each of the three authors. Drawn as a polar coordinate system, where the radial coordinate approximates network distance to the focal author in the center, and differences in the angular coordinate resemble community structure \cite{schulz2018visualizing}. Citing authors are marked in purple. }
\label{sindexexamples}
\end{figure}

 Using the \textit{network potential}, we can quantify 1) the value of an author's network position and 2) how their performance deviates from this baseline. Specifically, the $s$-index ("social" index) assigns an author coordinates in a 2-d space ($s_{\operatorname{pos}}$, $s_{\operatorname{pers}}$) (see Fig.~\ref{sindexexamples}).   The \textit{position} $s$-index ($s_{\operatorname{pos}}$) is the percentile rank of an author's \textit{network potential}, within the whole population. The \textit{personal} $s$-index ($s_{\operatorname{pers}}$) is the percentile rank of the empirical citations an author accrues within the group of 1,000 authors with the most similar network positions.  

\subsection*{Interpretation of the $s$-index}

 Now that we have established the strong correlation between network position and citation accrual, we investigate how the effect of network position on citations relates to other author characteristics. One potential explanation for the link between network position and citation accrual is that position is correlated with other    measures of a scientist's career success or ability, which actually drive the citations. We consider four different proxies for conventional success metrics that we can extract from our data: (1) the number of articles an author has published (productivity), (2) the number of citations they accrue over their career (i.e. career success), (3) the number of collaborators they have published with (a measure of how well-connected they are, which is simply the degree in the co-author network), and (4) career length. Simple linear correlation analysis reveals that these four metrics are clearly correlated with \textit{network potential} but they capture different variability in citation accrual between authors (see SI).     

 The system we are studying is complex, with multiple sources of feedback, spillovers and endogeneity. Thus, definitively establishing causality through an observational study is not possible. However, we can control for the effect of these four measures of success, as well as other potential confounds through a matched observational experiment. Specifically, using multiple observable covariates, we match each author that experiences a change in network position to another that does not (see Materials and Methods for more information on the matching procedure). We find that the first group experiences an increase in citations accrued in the next year that is significantly larger (see Fig.~\ref{volumeind} D). Interestingly, this effect is almost as strong as the effect calculated without comparing to a matched sample. Thus, we conclude that the effect of network position on citation accrual is robust to confounding. However, since multiple conditions for causal inference are not met (most importantly due to feedback and network spillovers), this cannot necessarily be interpreted as the treatment effect.   

\begin{figure}
\begin{adjustwidth}{-6.0cm}{-0mm}
\includegraphics[width=18.0cm]{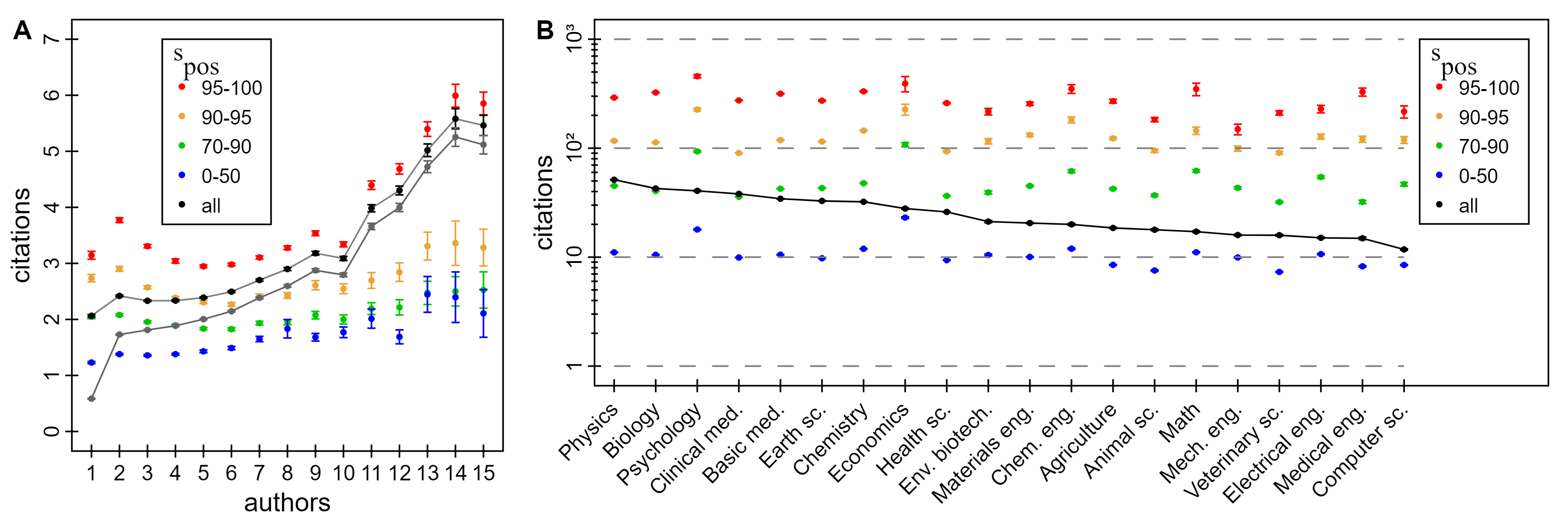}
\end{adjustwidth}
\caption[Can network position account for differences in citation impact across multi-author publications and scientific disciplines?]{\textbf{Can network position account for differences in citation impact across multi-author publications and scientific disciplines?} \textbf{(A)} Citations received by papers within the first two years of publication in 2010, as a function of the number of authors. We consider statistics over all of the analyzed authors (black, and for comparison, statistics from all publication data in gray) and also partitioned according to the $s_{\operatorname{pos}}$ of authors (specifically the author with the maximal $s_{\operatorname{pos}}$). Although there is a clear increase in citations with the number of authors in the first case, we can see that this effect is purely driven by authors in the highest $s_{\operatorname{pos}}$ class (red). In all other $s_{\operatorname{pos}}$ classes average citations stay constant independent of the number of authors. \textbf{(B)} Average number of citations authors of the 20 largest fields received in 2010, partitioning authors by different $s_{\operatorname{pos}}$ classes as in (A). Variation in citation across fields is reduced when we control for the network potential in this way, but there is still some variation due to field differences.}
\label{multiauthor}
\end{figure}

\subsection*{Applications of the $s$-index}

 The $s$-index can shed light on open questions regarding the comparative impact of (1) publications with different numbers of authors and (2) authors in different disciplines. In both of these cases the network effects vary together with the comparison groups. We make an impact comparison with fixed $s_{\operatorname{pos}}$ to control for these network differences.       

\subsubsection*{Evaluating the impact of multi-author collaborations}

Research shows that there has been a profound increase in the degree of collaboration among researchers \cite{borner2010multi}. There is evidence that this is a beneficial trend for science. For instance, publications resulting from team work receive more citations, on average \cite{wuchty2007increasing}. One hypothesis explaining the greater success of team work is that the work produced is of higher merit since many authors can tackle problems too complex or too laborious to solve individually. Here, we test to what extent this difference in impact can be accounted for by the network position of the authors of a publication. Concretely, we compare the citations received within two years of publication (results are robust to different time-windows) by publications with a different number of authors, but hold the maximum $s_{\operatorname{pos}}$ of the authors fixed (see Fig.~\ref{multiauthor} A). When we consider all publications we find the expected increase in citation impact with number of authors. However, when we look more closely, this effect seems to be restricted to an elite segment of publications. Specifically, when all the authors on a publication have a $s_{\operatorname{pos}}$ below 95, there is no significant difference in citation impact with different author team sizes (if we control for $s_{\operatorname{pos}}$). However, papers with at least one author of $s_{\operatorname{pos}}$ greater than 95 do exhibit a marked increase in citation impact as the number of authors increases above ten. Interestingly, even for the well-connected authors, performance is relatively flat for teams smaller than ten, except for an unexpectedly high citation count for papers with two or three authors.   

\subsubsection*{Comparing impact across disciplines}

Comparing the success of authors working in different fields is challenging. Citations are very susceptible to differences in conventions and the size of the fields. Interestingly, \cite{radicchi2008universality} shows that for a number of fields citation distributions follow universal behavior if rescaled by the average number of citations a field receives. However, to effectively use such a normalization for comparison between fields, boundaries between them need to be defined, and unfortunately, discipline classifications typically give only a rough approximation of the real community structure. Also, equalizing all differences in citation volume between fields obscures variation that is due to substantive differences in impact.

 The $s$-index can be used to make inter-field comparisons, using network information to control for many of the differences in the size and density of scientific communities, without requiring the explicit delineation of fields.    For example, we can see in Fig.~\ref{multiauthor} B that citation impact varies strongly between authors working in different scientific fields but this variation is greatly reduced if we compare authors with similar $s_{\operatorname{pos}}$. Furthermore, if we consider the 23 sub categories of the OECD science and technology classification that had at least 10,000 authors in 2010, the mean, standard deviation, and coefficient of variation of citations counts are $(\mu, \sigma, \frac{\sigma}{\mu})=(25.74, 10.89, 0.42)$, while the $s_{\operatorname{pers}}$ has corresponding values of $(48.66, 6.35, 0.13)$.   This indicates that much, but not all, of variation in performance across fields can be explained through network differences alone.   It is important to note, however, that strong differences persist in the performance of some fields even when we control for network potential. This is likely due to different co-authorship, publication and citation conventions. We can refine the $s$-index to factor in such field-specific differences by adding explicit information that limits comparisons to scientists working in similar fields.   For example, if we only compare authors working in the same labeled field, performance becomes more regular across fields (see SI for more details). Such a method can be used to provide a more equal comparison of scientists across fields.  

\section*{Discussion}

We have introduced a novel way of conceptualizing and quantifying a scientist's network position and used it to determine the systematic relationship between network position and citation accrual. We find that network position is a good predictor of citation accrual. Furthermore, although position is correlated with other, commonly considered factors of scientific success, it captures information on citation accrual that these do not. To control for the effect of these other factors, we match authors using observable covariates in our dataset and compare the change in citations experienced by   individuals that change their network position (i.e. a treatment group) to those that experience no change in network position (control group). We find an increase in citations in the treatment group that is significantly different from no treatment. The latter experience close to no change in citation accrual.   Thus we can conclude that the effect of network position on citations is not a spurious correlation due to observed sources of variability between scientists. Nonetheless, due to endogeneity and spillovers intrinsic to this system, a baseline that captures a pure network effect may not be possible to achieve and in any case requires a dataset with richer author metadata. This is an important avenue for future research. 

 Using the predictive power of network position we estimate the citation potential of an author's network position. We then construct an indicator, the $s$-index, that ranks (1) the citation potential of an author's network, compared to all scientists, and (2) their realized citations, compared to scientists with a similar network position. The $s$-index can be used to better quantify performance, assessing for instance who has built a network that has the potential to spread an important idea, or revealing hidden scientists who are performing above what is expected based on network position. This new index makes a significant contribution because it: (1) extracts information from a novel data source, (2) explicitly measures previously confounded components of performance, (3) has a basis in the theory of information diffusion on networks, and (4) is transparent.    

 Beyond comparing individual performance, our general approach and the $s$-index allow us to investigate the role of network differences in scientific performance of author teams of different sizes and of different fields. This is especially important as trans-disciplinary work and large teams become more common in science. Our results indicate that systematic differences in network position account for many differences in performance, and therefore we must be careful in how we interpret these. However, some differences cannot be predicted using network information, posing interesting questions for future research. 

 Our approach can easily be extended to additional indicators that characterize citation dynamics relative to scientific interpersonal networks. For example, we can introduce citation measures to quantify how citation sources are distributed throughout the network using the mean or entropy of an author's citation distance distribution. Furthermore, here we have only discussed a static picture, but the temporal evolution of an author's network position and of their $s$-index provides a fuller and novel perspective on scientific careers.   

 There are many different mechanisms that could bring about the correlation we find between network position and citations accrued. Self-citations and citations of co-authors may be strategically used (and misused) for self-promotion or to promote favorable colleagues. However, the network effects we discuss do not require this, and may be simply due to real differences in the quality of scientific work, the variation of publication conventions and size of different scientific communities, or the dynamics of information diffusion.   We currently focus on the co-authorship network due to the extensive documentation available over disciplines and time. However, applying our approach to alternate or multiple networks of interpersonal scientific interactions could help to disentangle the different mechanisms driving citation. Interesting examples are networks of acknowledgments, Twitter networks, or face-to-face interactions in conferences.

\section*{Materials and Methods}

\subsection*{Co-authorship network}

 This large-scale empirical study analyzes network data from over 13 million scientific careers with at least two publications, which are extracted from Clarivate Analytics’ Web of Science covering the period 1950-2015 and a wide range of scientific disciplines. About 54 million publication records linked by 729 million citations were name disambiguated using a method specifically designed for this database \cite{schulz2014exploiting}. Additionally, we replicated all experiments on the complete Microsoft Academic graph, which produced consistent results (see SI for more details). We consider researchers who have published at least once in the last five years, and have a minimum of two publications during their career.   The network is unweighted, and a link exists between a pair of authors if they have collaborated in the past. Experiments with different weighting schemes (weighted by the number of collaborations and decreasing weights over time) did not reveal significantly deviating results on a statistical level, and thus unweighted links were preferred for simplicity and computational efficiency. Since distances can only be computed for connected network components, we only consider authors in the largest component.

\subsection*{Distance computations}

Geodesic distances between authors were determined by computing the shortest path between all author pairs using a breadth-first search algorithm, posing a computational complexity of $\Omega(n+e)$, where $n$ is the number of active authors in a year (up to about 5 million for most recent years) and $e$ the number of pair-wise co-authorships. Computing the index for an individual researcher requires a pre-computation of distances of all authors in the database. We parallelize the computation by author and year, so that the 3.4$*10^{14}$ distances can be computed on a 1,000 core cluster in a couple of days.   Since citations and co-authorship events are only known with yearly timestamps, we compute distances at year $y$ by taking the average of the distance on the co-author network of the previous year $A(y-1)$ and the current year, $A(y)$.

\subsection*{$k$-nearest neighbor regression}

\begin{figure}
\begin{adjustwidth}{-6.0cm}{-0mm}
\includegraphics[width=18.0cm]{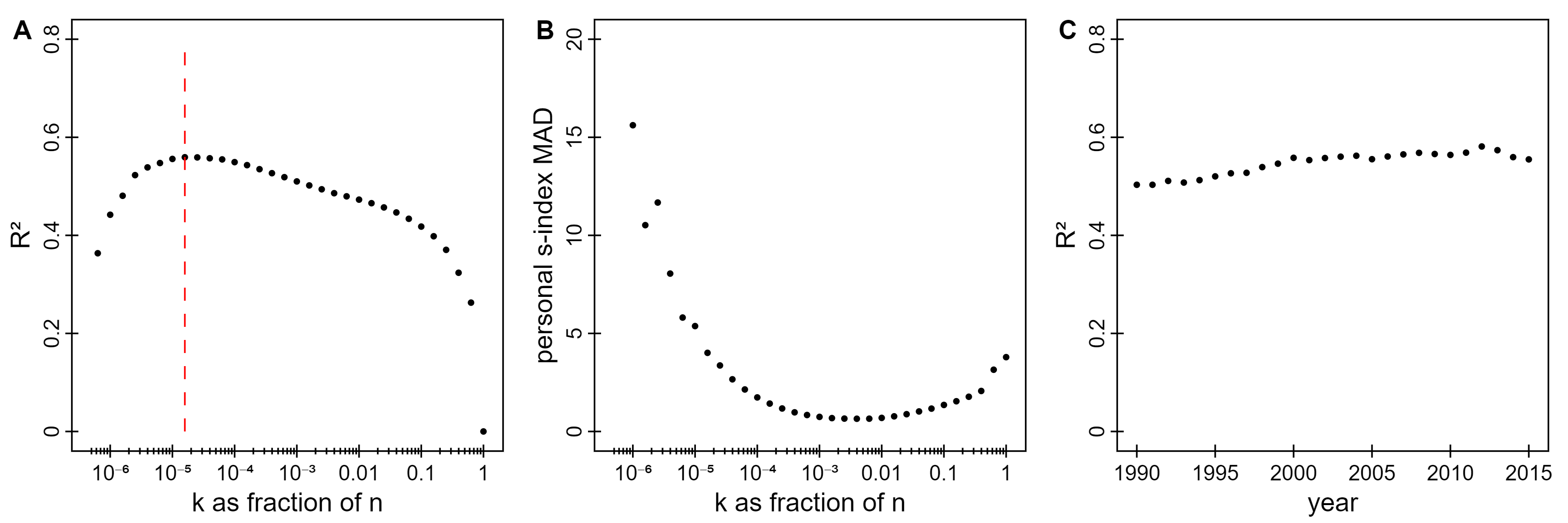}
\end{adjustwidth}
\caption[Choosing $k$ parameter for nearest neighbors in network potential model]{\textbf{Choosing $k$ parameter for nearest neighbors in network potential model.}  Applying a 10-fold cross validation with all authors in 2010, we find the optimal $k$ by \textbf{(A)} maximizing the $R^2$ value of the \textit{network potential} model and \textbf{(B)} minimize the mean absolute difference in $s_{\operatorname{pers}}$ scores when increasing k as fraction of n by 0.2 in log space. $k$ is then chosen as a compromise between predictability (high agreement between actual and expected citations) and stability (low deviation in resulting $s_{\operatorname{pers}}$ ranking when changing $k$). \textbf{(C)} Predictive power computed for each year using $k=$1,000.}
\label{choosingk}
\end{figure}

For each author with at least 1 citation, we find other authors with a similar distance distribution $P_i(d)$ using a $k$-nearest neighbor (k-NN) regression \cite{altman1992introduction} with $k=$1,000 authors. We choose $k$ to achieve an optimal bias-variance trade-off and individual ranking stability. To do this, we maximize the $R^2$ value of the \textit{network potential} model through a 10-fold cross validation with authors in 2010 (see SI Fig.~\ref{choosingk}). Additionally, we measure individual deviations in resulting ranking values depending on the choice of $k$. Distributions are compared by measuring the Euclidean distance between the 10-dimensional vectors with elements $P_i(d=x)$, $x\in [0..9]$, standardized with $\mu=0$ and $\sigma=1$.   

\subsection*{Matched comparison}

To reduce confounding effects, we establish a control group of authors with a non-increasing number of citations (or network potential) and match most similar authors according to the observable variables.   We use a 1-nearest neighbor matching method without replacement and a distance measured in Euclidean space formed by the standardized variables ($\mu=0$, $\sigma=1$) results in a good balance of the distributions of all variables in treatment and control groups, measured by the (average) standardized differences of the means of treatment and control (network potential 0.012, number of citations 0.011, career age 0.014, total number of papers 0.008, total number of collaborators 0.013 and total number of citations 0.021). A Wilcoxon signed-rank test shows that the chance that the differences between treatment and control sample pairs follow a symmetric distribution is negligible (a vanishing p-value).

\bibliography{literature}
\clearpage
\beginsupplement

\section*{Supporting information}

\subsection*{Construction of the network-driven citation model and the $s$-index}

As discussed in the main text, our \textit{network potential} is constructed using a $k$-nearest neighbor regression model, i.e. an average of most similar neighbors. Here, we compare the $k$-NN approach with alternative models. All models have in common that they only use information about the collaboration network:

\begin{enumerate}
\item Average distance $\langle d \rangle$: From $P(d)$ we can compute a single mean distance value that describes how far away an author is on average from any publication. We can see in Fig.~\ref{neteffect} F that there is a strong relationship between average distance and the number of received citations. We fit an exponential function $log10(c) = p_1 \exp(-p_2(\langle d \rangle - d_{min}))+p_3$.
\item Estimating a global $P(c|d)$, i.e., the probability of a citation given an article at a distance $d$. This is an average model, evaluating each distance individually. The average can be computed from $P(d|c)$ and $P(d)$ by counting both the number of articles and the number of citations over all authors at each distance.
\item The average score of the set of authors with most similar $P(d)$ compared to the focal author, using $k$-NN. The \textit{network potential} can be directly computed for each author, without the need for an intermediate model.
\end{enumerate}

Choosing the third approach has qualitative and quantitative advantages. For model (1), different shapes of $P(d)$ can lead to the same average distance. Additionally, despite the good fit, we have no theoretical argument for selecting an exponential function as a model. In (2) we assume a linear relationship between the number of publications and citations at a certain distance, ignoring potentially different dynamics of networks with varying density. (3) is essentially model-free and relies on a large number of observations and the high regularity of the relationship between collaboration network and citations. We test all three models using a ten-fold cross-validation on all authors in 2010. The computed \textit{network potential} of each model are used as a predictor for the actual number of citations. Linear regressions result in an $R^2$ of $0.337$, $0.45$, and $0.55$, respectively, which again makes the $k$-NN approach the best one.

\subsection*{Role of career vintage, productivity, collaborators, and overall impact}

\begin{figure}
\begin{adjustwidth}{0.0cm}{-0mm}
\includegraphics[width=12.0cm]{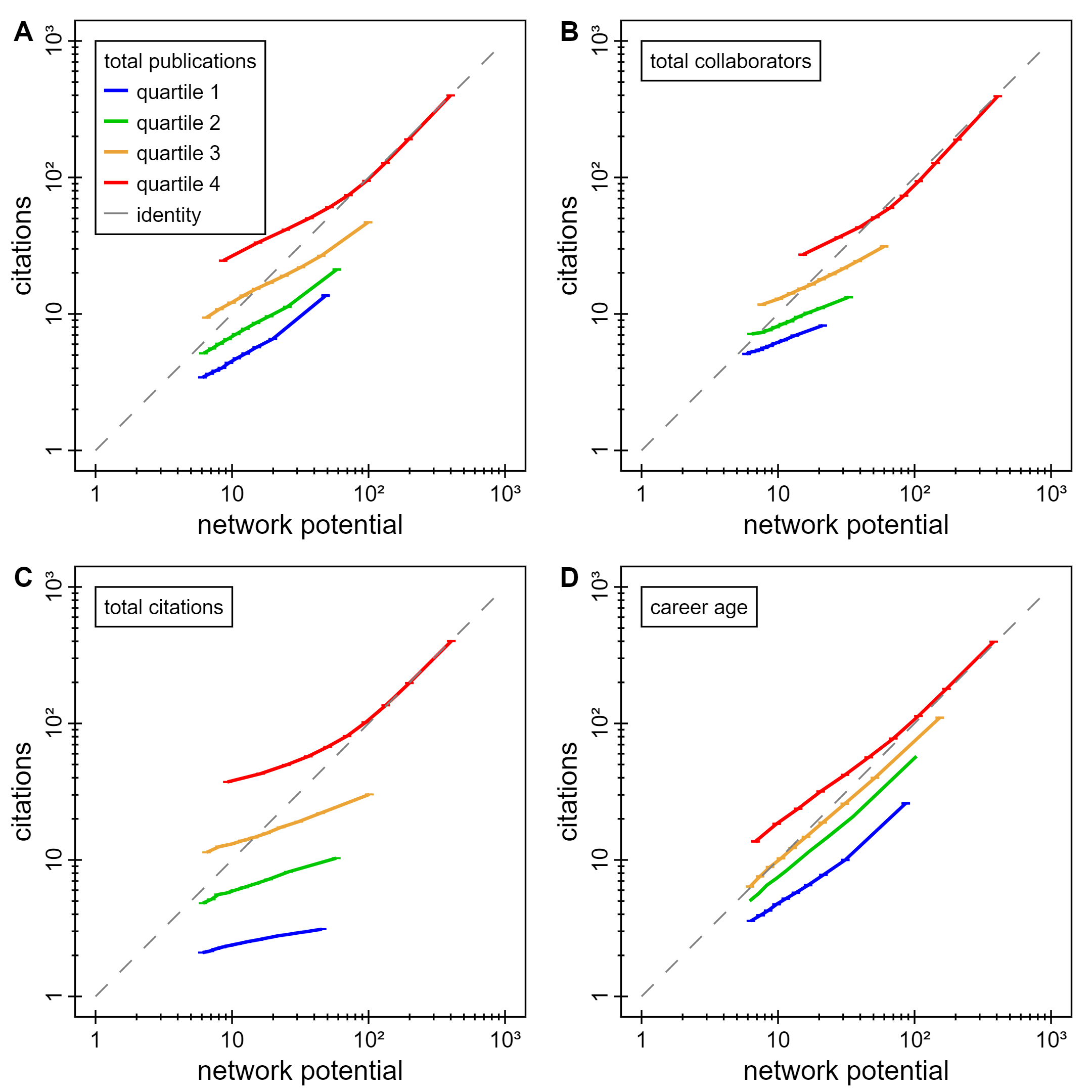}
\end{adjustwidth}
\caption[Comparing network effects with career age and traditional performance measures]{\textbf{Comparing network effects with career age and traditional performance measures.}  (Data: all authors active in 2010, citations received in 2010.) Average citation volume during a specific year, as a function of the \textit{network potential}. We partition researchers into different quartiles of performance according to other standard metrics. Each panel displays a different metric: \textbf{(A)} the number of publications, \textbf{(B)} the number of previous collaborators, and \textbf{(C)} the total number of citations they receive in their career, and \textbf{(D)} the number of years since the first publication (career age). The dotted line marks the identity, where the network position fully accounts for citations. Above this line authors are over-performing relative to their \textit{network potential}, and below it they underperform. Although network position captures a lot of the variation in citation counts, the other measures of performance capture different sources of variation, driven by factors such as reputation, experience, or differences in the quality of work of individual researchers.}
\label{factors}
\end{figure}

How are common measures of success correlated with both the \textit{network potential} and the citation impact of an author? Starting with productivity, if citations were random, we would expect the number of published articles to explain all variance in citation counts. As can be seen in Fig.~\ref{factors} A, although productivity does correlate with citations (Pearson correlation coefficient of 0.66), the network position seems to have a stronger effect on the ability to receive citations (correlation coefficient of 0.72). In Fig.~\ref{factors} B we investigate how the total number of different previous collaborators affects citations. This total number of collaborators is the first-order effect of a high level of collaboration. It is trivial to compute but again is less correlated with citations than the network position which takes higher-order effects into account (correlation coefficient of 0.42 vs. 0.72). In other words, while we observe a higher expected citation volume with more co-authors, for outstanding success it is also essential that co-authors are well-connected themselves, thereby shrinking the network distances. Career success (total citations) is clearly a powerful predictor of citations in a given year. On average, an author that receives many citations over their career performs at this level in the chosen year, independent of their network position (see Fig.~\ref{factors} C). Surprisingly however, total citations are only slightly more correlated with current citations than the \textit{network potential} (correlation coefficient of 0.77 vs. 0.72).

 Lastly, while the number of years since the first publication (career age) is fairly predictive for citation counts (correlation coefficient of 0.40), the relationship between network position and citations remains robust for different levels of career age (see Fig.~\ref{factors} D).

\begin{figure}
\begin{adjustwidth}{-6.0cm}{-0mm}
\includegraphics[width=18.0cm]{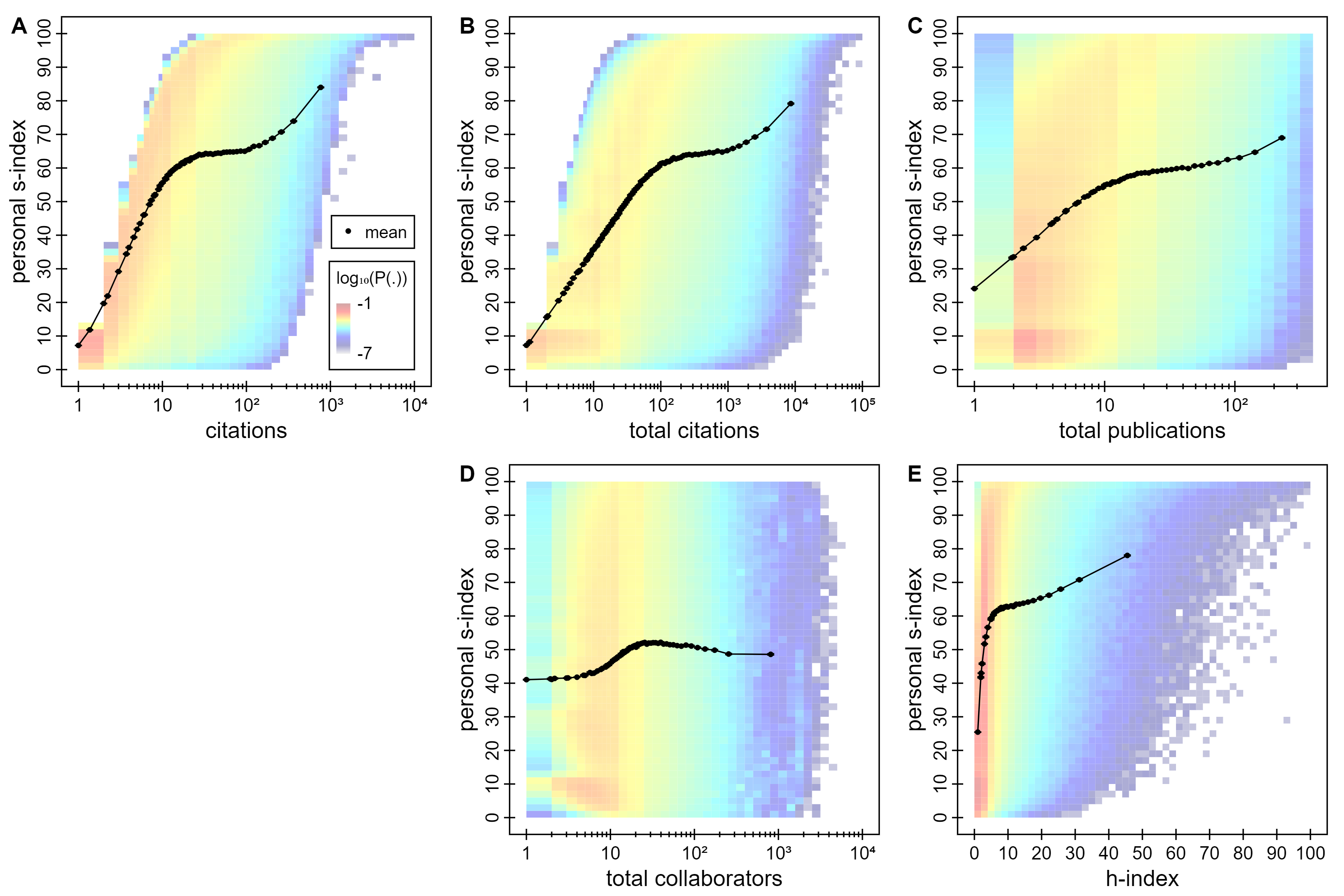}
\end{adjustwidth}
\caption[Comparison to other bibliometric indices]{\textbf{Comparison to other bibliometric indices.} \textbf{(A)} Density plot for authors with number of citations received in 2010 and corresponding $s_{\operatorname{pers}}$. Citation distributions are usually skewed with many authors receiving only one citation and few authors getting many times more citations than average. We observe a lower bound and upper bound for the $s_{\operatorname{pers}}$ given a certain citation count. \textbf{(B)} Previous citation success and, to some extent, \textbf{(C)} productivity are predictive for success in subsequent years, and therefore also for the $s$-index, albeit with great variance. \textbf{(D)} At all degrees of collaboration, any $s_{\operatorname{pers}}$ value is possible. \textbf{(E)} While very high $h$-indices correspond to high $s$-indices, $s_{\operatorname{pers}}$ captures more information about very small $h$-index values and in moderate $h$-index ranges the two indices capture different information.}
\label{sindexOtherIndices}
\end{figure}

 Alternatively, can the conventional measures of success account for an author's deviation from their \textit{network potential}, i.e. their $s_{\operatorname{pers}}$? We can see that the variation in these measures does explain some of the variation in the relationship between the actual citation count and the \textit{network potential}. This relationship is precisely $s_{\operatorname{pers}}$. For example, we can observe that authors in the highest performing quartile, according to each one of the measures of success that we consider, perform (on average) at least as well as expected due to the network alone. In all other quartiles, authors underperform on average, except for a few cases where the expected citations due to network position are very low. However, the correlation between $s_{\operatorname{pers}}$ and these other measures is small (correlation coefficient of 0.20 for productivity, 0.03 for collaborators and 0.24 for past citation). Thus, there are other important drivers of idiosyncratic deviations in performance. Fig.~\ref{sindexOtherIndices} provides a comparison of $s_{\operatorname{pers}}$ with other common bibliometric indicators. 

\begin{figure}
\begin{adjustwidth}{6.0cm}{-0mm}
\includegraphics[width=6.0cm]{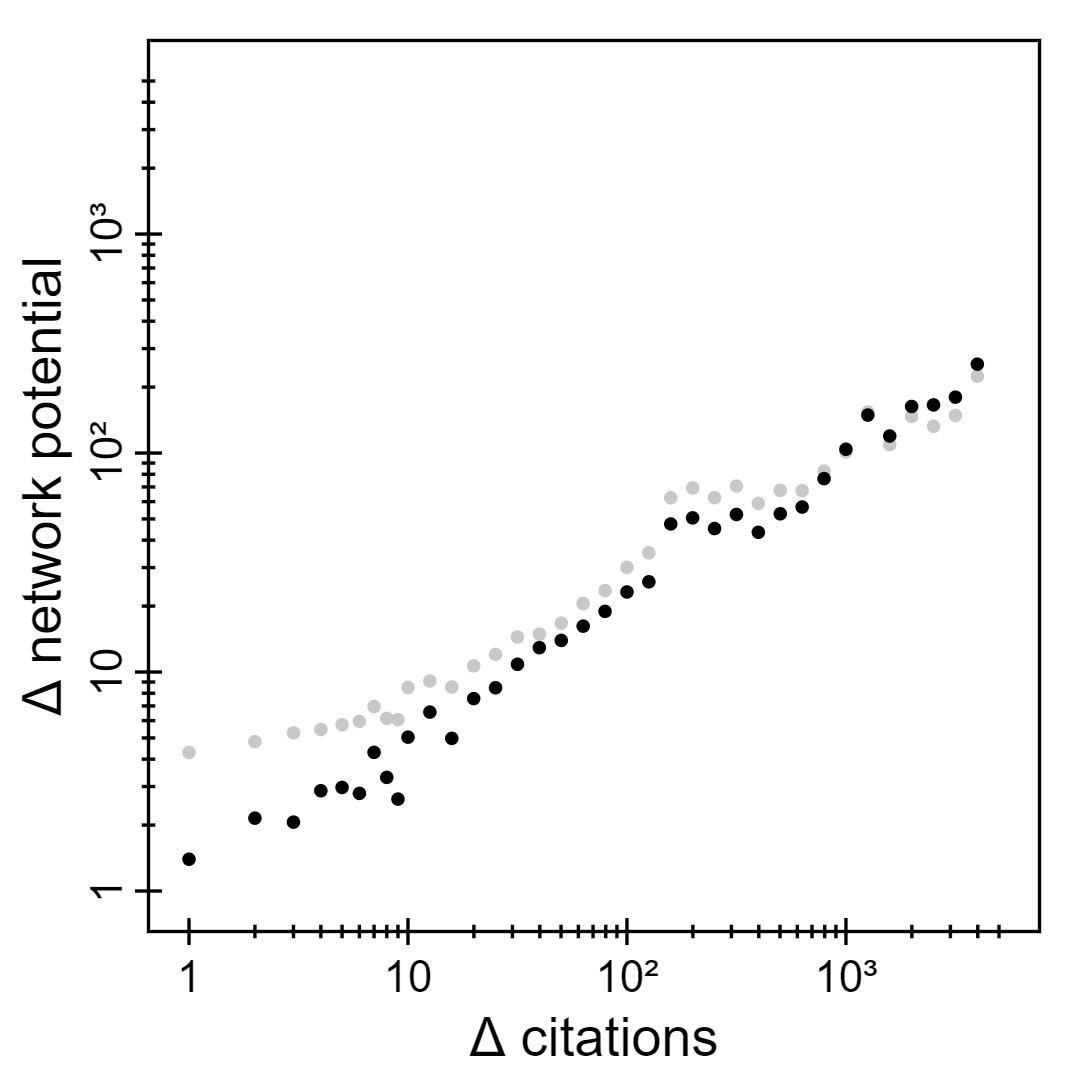}
\end{adjustwidth}
\caption[Effect of an increase in network potential or citation success]{\textbf{Effect of an increase in network potential or citation success.}  Data: all authors in 2011 who experienced an increase in citations compared to their value in 2010 (See also Fig.~\ref{volumeind} D for an increase in network potential). Log-binned by magnitude of change. For each bin, we compute the average difference of network potential (or citations, respectively) received in 2012 compared to 2011. The plot displays the treatment outcome in grey, and the treatment subtracting the matched control group in black.   Compared to Fig.~\ref{volumeind} D, we can conclude that while both variables typically increase over the course of a career, the immediate gain in network potential followed by an increase of citations is weaker than the gain in citations followed by an increase in network potential. }
\label{matching}
\end{figure}

\subsection*{Disciplines}

To what extent does the $s$-index normalize differences in citation counts across scientific disciplines? Table \ref{communitiesTable} informs about the variance of field-specific mean citation counts and mean s-indices. By definition, the percentile values of the $s$-index generate a uniform distribution between 0 and 100 with a mean of 50. If network effects predicted all field differences, then we would also observe a uniform $s_{\operatorname{pers}}$ distribution for each individual field (see the Global k-NN columns in the table). However, although variation is reduced between fields, we do not observe a uniform distribution. Math, for example, a field with low average citation counts and less well-connected authors due to the comparatively lower degree of collaboration and productivity, tends to have lower $s_{\operatorname{pos}}$ scores and thus higher $s_{\operatorname{pers}}$ scores than the absolute number of citations would suggest. In the case of Computer Science, where low citation counts could be attributed to an incomplete database coverage due to missing conference proceedings, the $s$-index also compensates for lower connectivity, resulting in an average $s_{\operatorname{pers}}$ of around 50 as well. In some disciplines, especially in the Social Sciences, authors can attract more citations than the network potential predicts, generating a greater number of high $s_{\operatorname{pers}}$ values. These authors are frequently compared to authors from other fields that are more common in the database (e.g. Natural Sciences and Medicine). An extension of the $s$-index (refered to as field k-NN in the table) explicitly controls for between-field differences explicitly limits the k-NN regression to authors within the same field. In this case we necessarily end up in almost uniform distributions of $s_{\operatorname{pers}}$ for each field.  

\begin{table}
\caption[Average number of citations and $s$-index by discipline]{\textbf{Average number of citations and $s$-index by discipline.}  Data: all authors active in 2010. OECD sub-disciplines with more than 10,000 authors, and no fields labeled with "Other...". Mean number of citations an author receives in 2010. A global $k$-NN (as used for all results in the main text) selects authors that have a similar network profile from the set of all authors in the database. In contrast, the within-field $s$-index only compares authors in the same OECD subfield. The global version already significantly decreases field differences (coefficient of variation of field means reduces from 0.42 to 0.13 for $s_{\operatorname{pers}}$). Remaining variance could be due to field-specific citation and collaboration practices, actual differences in citation utility, or simply varying database coverage.}
\label{communitiesTable}
\begin{adjustwidth}{-6cm}{-0mm}
\resizebox{18cm}{!}{%
\begin{tabular}{ llrlllll }
&   &  &  &  Global k-NN & & Field k-NN & \\
\hline
OECD main field & OECD sub field &  Authors & Citations & $s_{\operatorname{pos}}$ & $s_{\operatorname{pers}}$ & $s_{\operatorname{pos}}$ & $s_{\operatorname{pers}}$  \\
\hline
Natural sc. & Math & 35,363 & 17.14 & 26.14 & 53.23 & 37.55 & 52.11  \\
 & Computer sc. & 26,693 & 11.76 & 24.34 & 47.68 & 26.56 & 51.78 \\
 & Physics & 239,435 & 51.33 & 55.38 & 46.93 & 56.09 & 50.66 \\
 & Chemistry & 244,734 & 32.21 & 48.77 & 48.93 & 54.14 & 50.39 \\
 & Earth sc. & 121,460 & 32.80 & 50.25 & 46.81 & 46.99 & 50.89 \\
 & Biology & 458,232 & 42.66 & 57.42 & 44.48 & 54.40 & 50.13 \\
 Engineering &	Electrical eng. & 29'762 & 15.05 & 25.67 & 53.52 & 37.58 & 51.77  \\
 & Mech. eng. & 26,699 & 15.95 & 31.10 & 49.93 & 37.29 & 51.45 \\
 & Chem. eng. & 13,132 & 19.99 & 32.15 & 55.02 & 47.02 & 51.62 \\
 & Materials eng. & 43,297 & 20.54 & 37.67 & 49.57 & 41.35 & 51.19 \\
 & Medical eng. & 12,100 & 14.89 & 40.03 & 41.76 & 30.69 & 51.94 \\
 & Env. biotech. & 11,762 & 21.19 & 45.43 & 45.05 & 45.94 & 51.60 \\
 Medical \& health sc. & Basic med. & 230,214 & 34.36 & 51.62 & 45.55 & 50.25 & 50.32  \\
 & Clinical med. & 718,048 & 38.06 & 52.61 & 43.49 & 49.39 & 50.52 \\
 & Health sc. & 99,009 & 26.00 & 45.66 & 45.27 & 42.27 & 50.69 \\
 Agricultural sc. & Agriculture & 31,824 & 18.53 & 40.94 & 45.41 & 34.40 & 51.41  \\
 & Animal sc. & 12,718 & 17.84 & 42.62 & 41.49 & 32.54 & 51.67 \\
 & Veterinary sc. & 20,233 & 15.89 & 40.93 & 41.49 & 29.53 & 51.11 \\
 Social sc. & Psychology & 32,879 & 40.61 & 37.26 & 61.16 & 58.72 & 50.99 \\
 & Economics & 23,168 & 27.92 & 21.22 & 66.46 & 61.87 & 50.57 \\
\hline
 & & $\mu$ & 25.74 & 40.36 & 48.66 & 43.73 & 51.14  \\
 & & $\sigma$ & $\pm$10.89 & $\pm$10.57 & $\pm$06.35 & $\pm$10.08 & $\pm$00.58  \\
 & & $\frac{\sigma}{\mu}$ & 00.42 & 00.26 & 00.13 & 00.23 & 00.01  \\
\hline
\end{tabular}}
\end{adjustwidth}
\end{table}

\subsection*{Evolution of scientific careers}

\begin{figure}
\begin{adjustwidth}{-6.0cm}{-0mm}
\includegraphics[width=18.0cm]{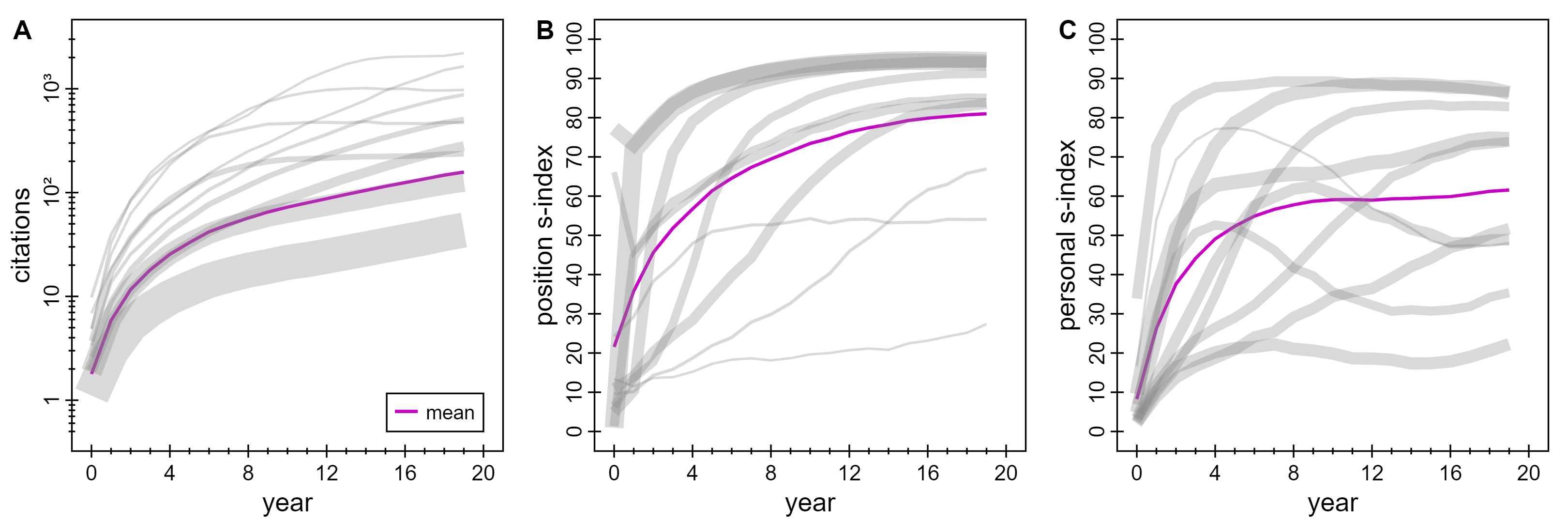}
\end{adjustwidth}
\caption[Temporal analysis]{\textbf{Temporal analysis.}  Data: 10,000 researchers who started their career in 1990 with at least 20 years of publication history. $k$-means clustering of curves with $k=10$. Line thickness is proportional to cluster size. The cyan line indicates the mean value over all authors in the sample. \textbf{(A)} Received citations in career year (non-cumulative). \textbf{(B)} $s_{\operatorname{pos}}$ by year. Some researchers already start at good network positions and can continue to improve their network potential. Here, data selection is biased towards more successful scientists, since we only consider researchers with a long productive career. \textbf{(C)} $s_{\operatorname{pers}}$ by year. While citation rates usually increase, $s_{\operatorname{pers}}$ can peak earlier in a career and continue with a downward trend.  }
\label{temporal}
\end{figure}

The time evolution of the $s$-index provides a different perspective on an author's career. The $s$-index is evaluated each year for all authors in the chosen database. Contrary to most indicators of scientific productivity and success, it can decrease with time. Some interesting patterns can be seen in Fig.~\ref{temporal}.   Characterizing scientific careers based on the dynamic behavior of their index, relative to the mean behavior could yield insight on the determinants of scientific success.

\subsection*{Replication with different data}

\begin{figure}
\begin{adjustwidth}{0.0cm}{-0mm}
\includegraphics[width=12.0cm]{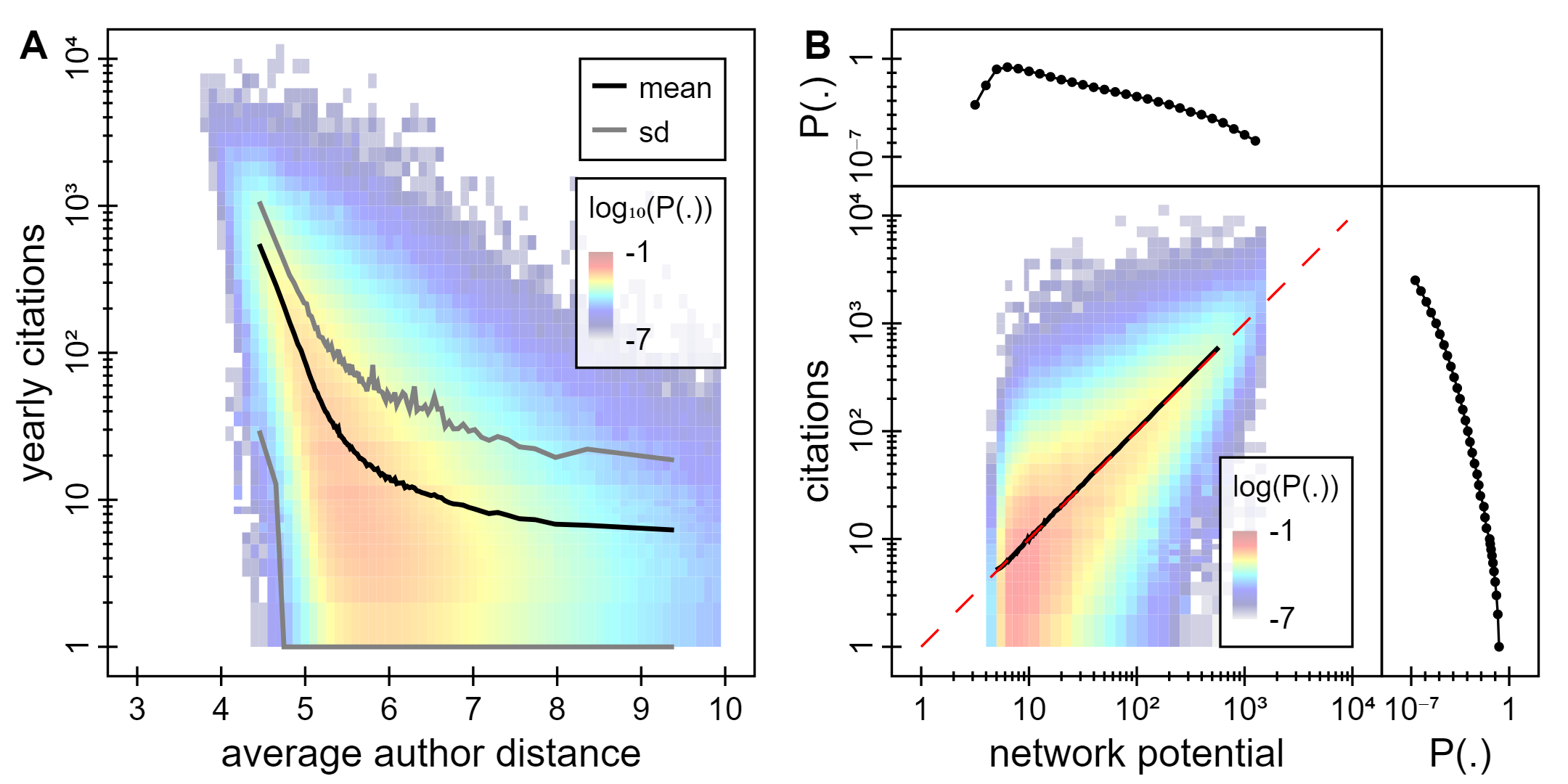}
\end{adjustwidth}
\caption[Experiment replication with Microsoft Academic Graph (MAG) data]{\textbf{Experiment replication with Microsoft Academic Graph (MAG) data.} \textbf{(A)} Citations an author receives in 2010, given their average author network distance. This plot is a replication of Fig.~\ref{neteffect} F. \textbf{(B)} Distributions of the network potential and the actual citations. Compare with Fig.~\ref{volumeind} C. }
\label{magdata}
\end{figure}

We successfully replicated all main results using the Microsoft Academic Graph (MAG) data. This dataset comprises of 166 million papers and 1.026 billion resolved citations and has different coverage than the WoS data, particulary for conference proceedings, that are the dominant type of publication in some fields such as Computer Science. Fig.~\ref{magdata} shows the results for the MAG data.

\end{document}